\documentclass[sigconf, anonymous=False]{acmart}

\usepackage{multirow}
\usepackage{float} 
\usepackage{amsmath}
\usepackage[ruled,linesnumbered]{algorithm2e}
\usepackage{graphicx}
\usepackage{subfigure}
\usepackage{epstopdf}
\usepackage{tablefootnote}
\usepackage[flushleft]{threeparttable}
\usepackage{diagbox}
\usepackage{booktabs}
\usepackage{siunitx}
\usepackage{footmisc}
\usepackage{footnote}
\usepackage{enumitem}
\usepackage[normalem]{ulem}
\useunder{\uline}{\ul}{}

\copyrightyear{2023}
\acmYear{2023}
\setcopyright{acmlicensed}
\acmConference[SIGIR '23] {Proceedings of the 46th International ACM SIGIR Conference on Research and Development in Information Retrieval}{July 23--27, 2023}{Taipei, Taiwan.}
\acmBooktitle{Proceedings of the 46th International ACM SIGIR Conference on Research and Development in Information Retrieval (SIGIR '23), July 23--27, 2023, Taipei, Taiwan}
\begin{document}

\title{SAILER: Structure-aware Pre-trained Language Model for Legal Case Retrieval}

\author{Haitao Li}
\affiliation{DCST, Tsinghua University}
\affiliation{Quan Cheng Laboratory}
\email{liht22@mails.tsinghua.edu.cn}

\author{Qingyao Ai}
\affiliation{DCST, Tsinghua University}
\affiliation{Quan Cheng Laboratory}
\email{aiqy@tsinghua.edu.cn}

\author{Jia Chen}
\affiliation{DCST, Tsinghua University}
\affiliation{Quan Cheng Laboratory}
\email{chenjia0831@gmail.com}

\author{Qian Dong}
\affiliation{DCST, Tsinghua University}
\affiliation{Quan Cheng Laboratory}
\email{dq22@mails.tsinghua.edu.cn}

\author{Yueyue Wu}
\affiliation{DCST, Tsinghua University}
\affiliation{Quan Cheng Laboratory}
\email{wuyueyue@mail.tsinghua.edu.cn}

\author{Yiqun Liu}
\affiliation{DCST, Tsinghua University}
\affiliation{Zhongguancun Laboratory}
\email{yiqunliu@tsinghua.edu.cn}

\author{Chong Chen}
\affiliation{Huawei Cloud BU}
\email{chenchong55@huawei.com}

\author{Qi Tian}
\authornote{Corresponding author}
\affiliation{Huawei Cloud BU}
\email{tian.qi1@huawei.com}

\renewcommand{\shortauthors}{Li, et al.}

\begin{abstract}
Legal case retrieval, which aims to find relevant cases for a query case, plays a core role in the intelligent legal system. Despite the success that pre-training has achieved in ad-hoc retrieval tasks, effective pre-training strategies for legal case retrieval remain to be explored. Compared with general documents, legal case documents are typically long text sequences with intrinsic logical structures.
However, most existing language models have difficulty understanding the long-distance dependencies between different structures. 
Moreover, in contrast to the general retrieval, the relevance in the legal domain is sensitive to key legal elements. Even subtle differences in key legal elements can significantly affect the judgement of relevance. However, existing pre-trained language models designed for general purposes have not been equipped to handle legal elements.

To address these issues, in this paper, we propose \textbf{SAILER}, a new \textbf{S}tructure-\textbf{A}ware pre-tra\textbf{I}ned language model for \textbf{LE}gal case \textbf{R}etrieval. It is highlighted in the following three aspects: (1) SAILER fully utilizes the structural information contained in legal case documents and pays more attention to key legal elements, similar to how legal experts browse legal case documents. (2) SAILER employs an asymmetric encoder-decoder architecture to integrate several different pre-training objectives. In this way, rich semantic information across tasks is encoded into dense vectors. 
(3) SAILER has powerful discriminative ability, even without any legal annotation data. It can distinguish legal cases with different charges accurately. Extensive experiments over publicly available legal benchmarks demonstrate that our approach can significantly outperform previous state-of-the-art methods in legal case retrieval.

\end{abstract}




\maketitle

\section{Introduction}
Legal case retrieval serves as an important part of modern legal systems. For justice and fairness, it's necessary for legal experts to find and analyze relevant precedents to a specific case before making judgments. Recently, with the explosion of legal case documents, finding relevant ones among the whole corpus has become increasingly challenging. Therefore, studies on legal case retrieval systems have received considerable attention from both the legal and information retrieval communities~\cite{shao2020bert,bench2012history,yu2022explainable,ma2021lecard,althammer2021dossier}.

\begin{table}[t]
\caption{An example to introduce legal relevance. The two paragraphs are very similar in text, but legally irrelevant due to key legal elements.}
\label{table_case}
\begin{tabular}{p{0.95\columnwidth}}
\toprule
\textbf{Paragraphs A}: Person X, 24 years old, men, height 180 cm. In 2018, he entered the shopping mall five times where he \textbf{stole} one cell phone and two tablet computers, worth a total of 10,000,000. At five o'clock, he returned home and gave the above items to Person Y.     \\ \midrule
\textbf{Paragraphs B}: Person X, 24 years old, men, height 180 cm. In 2018, he entered the shopping mall five times where he \textbf{purchased} one cell phone and two tablet computers, worth a total of 10,000,000. At five o'clock, he returned home and gave the above items to Person Y. \\ \bottomrule
\end{tabular}
\vspace{-5mm}
\end{table}

In specific domains like legal case retrieval, constructing high-quality annotated datasets is often labor-intensive and prohibitive due to the need for domain expertise. However, state-of-the-art retrieval systems are usually constructed based on neural models trained with large-scale annotated data. Hence, IR researchers propose to utilize pre-trained language models (PLM), i.e., large-scale neural models trained without supervised data for language understanding, to conduct effective retrieval ~\cite{zhang2019ernie,yang2022thuir,floridi2020gpt,li2023towards}. Previous studies ~\cite{zhan2020repbert,karpukhin2020dense,chen2022axiomatically,dong2022incorporating} have shown that PLM such as BERT~\cite{devlin2018bert} and RoBERTa~\cite{liu2019roberta} significantly outperform existing neural retrieval models on passage and document retrieval datasets like MS MARCO and TREC DL in both zero-shot and few-shot settings~\cite{nguyen2016ms,craswell2020overview,xie2023t2ranking}.


Despite their superior performance in ad-hoc retrieval and open-domain search, the effectiveness of pre-trained language models has not been observed in legal case retrieval yet. Compared with traditional retrieval tasks, the application of PLMs to legal case retrieval presents two non-trivial challenges that have been overlooked by existing studies~\cite{shao2020bert}.

\begin{figure*}[ht]
  \centering
  \includegraphics[width=0.8\textwidth]{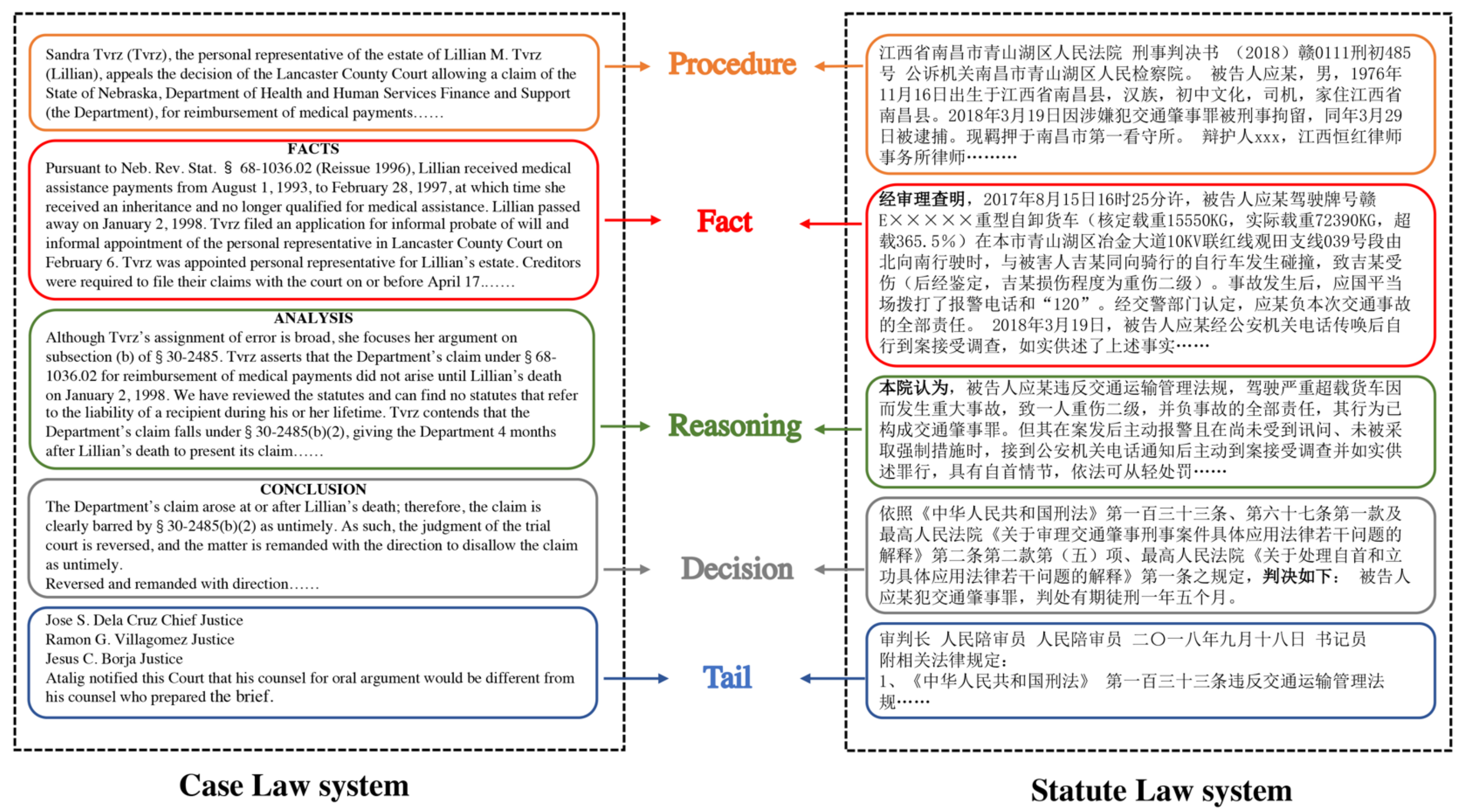}
  \vspace{-0.15in}
  \caption{Illustration of the legal case structure. The left case is from the United States (Case Law System) and the right case is from China (Statute Law System). The standard legal case documents can be organized into five parts: Procedure, Fact, Reasoning, Decision, and Tail.}
  \vspace{-5mm}
  \label{structure}
  \end{figure*}

\textbf{Challenge 1.} 
Legal case documents are typically long text sequences with intrinsic writing logic. As shown in Figure \ref{structure}, case documents in the Case Law system and Statute Law system \footnote{The two types of law systems that have been used in almost all countries in the world.} usually consist of five parts: Procedure, Fact, Reasoning, Decision, and Tail (we will discuss the details of these parts in Section 3). Each part represents a specific topic and varies from hundreds to thousands of words. These parts are written in standard legal logic and are usually correlated with each other. Existing PLMs either possess limited text modeling capacity, hindering their ability to model long documents~\cite{devlin2018bert,liu2019roberta}, or neglect the structures of legal case documents, preventing them from capturing the long-distance dependencies in legal writing logic~\cite{xiao2021lawformer}. Consequently, the performance of vanilla PLM-based retrievers is constrained.

\textbf{Challenge 2.} The concept of relevance in the legal domain is quite different from that in general search. In legal case retrieval, the relevance between two legal cases is sensitive to their key legal elements (e.g. ``forcibly took other people’s property", ``arbitrarily damaged other people’s belongings", etc.).  
Here, key legal elements include key circumstances and the legal concept abstraction of key circumstances or factors~\cite{ma2021lecard}. Cases without key legal elements or with different key legal elements may lead to different judgments.
For example, as shown in Table \ref{table_case}, the two paragraphs are usually considered relevant in ad-hoc retrieval because they share significant number of keywords and sentences. However, in legal case retrieval, these two paragraphs are irrelevant and could lead to completely different judgments due to the influence of key legal elements. Without guidance, open-domain PLM-based neural retrieval models have difficulty understanding key legal elements, resulting in suboptimal retrieval performance in the legal domain.

To tackle these challenges, we propose a \textbf{S}tructure-\textbf{A}ware pre-tra\textbf{I}ned language model for \textbf{LE}gal case \textbf{R}etrieval (SAILER). SAILER adopts an encoder-decoder architecture to explicitly model and captures the dependency between the Fact and other parts of a legal case document \textbf{(Challenge 1)}. Meanwhile, SAILER utilizes the legal knowledge in the Reasoning and Decision paragraphs to enhance the understanding of key legal elements \textbf{(Challenge 2)}.
Specifically, we encode the Fact paragraph into dense vectors using a deep encoder. Then, with help of the Fact vector, two shallow decoders are applied to reconstruct the aggressively-masked texts in the Reasoning and Desicions paragraphs respectively. In this way, SAILER takes full advantage of the logical relationships in legal case documents and the knowledge in different structures.

To verify the effectiveness of our method, we conducted extensive experiments on Chinese and English legal benchmarks in both zero-shot and fine-tuning settings. Empirical experimental results show that SAILER can achieve significant improvements over the state-of-the-art baselines.

We summarize the major contributions of the paper as follows: 

\begin{enumerate} 
\item We propose a novel pre-training framework for legal case retrieval, namely SAILER, which is the first work that exploits the structure of legal cases for pre-training.  
\item We propose several pre-training objectives to capture the long-distance text dependency between different structures and intrinsic writing logic knowledge by simulating the writing process of legal case documents. 
\item We conduct extensive experiments on public Chinese and English legal benchmarks. Experimental results demonstrate the benefits of modeling long-distance text dependency and utilizing structural knowledge in legal case retrieval\footnote{Code are available at \url{https://github.com/CSHaitao/SAILER}.}.
\end{enumerate}

\section{Related Work}
\subsection{Pre-trained Language Models for Dense Retrieval}

Dense retrieval (DR) typically uses a dual encoder to encode queries and documents separately and computes the relevance score by a simple similarity function (cosine or dot product). Many researchers have further improved the performance of DR by negative sampling and distillation~\cite{zhan2020repbert,qu2021rocketqa,fan2022pre,yang2023enhance,chenthuir}.

Researchers in the IR community have already begun to design pre-training methods tailored for DR~\cite{ma2022pre,wu2022contextual,wang2022simlm,lu2021less,gao2021unsupervised,liu2022retromae,gao2021condenser}. These approaches primarily aim to better represent contextual semantics with [CLS] embedding. They are based on the intuition that [CLS] embedding should encode important information in the given text to achieve robust matches. For example, Condenser ~\cite{gao2021condenser} and coCondenser ~\cite{gao2021unsupervised} design skip connections in the last layers of the encoder to force information aggregated into the [CLS] token. Recently, autoencoder-based pre-training has drawn much attention. The input sentences are encoded into embeddings to reconstruct the originals, forcing the encoder to generate better sentence representations. SEED-Encoder ~\cite{lu2021less} proposes to use a weak decoder for reconstruction. SIMIM ~\cite{wang2022simlm} and RetroMAE ~\cite{liu2022retromae} modify the decoding approach to strengthen the information bottleneck, which improves the quality of generated embeddings.

Despite the success, the autoencoder-based models cannot fully comprehend the logical relationship between the different structures in legal case documents, as they mainly rely on limited information in the corpus. Moreover, since the case text contains numerous facts that are not essential for judging the relevance of the cases, reconstructing the original text may decrease the discriminative power of dense vectors.

\subsection{Legal Case Retrieval}
There are two main categories of legal case retrieval models ~\cite{bench2012history}: Expert Knowledge-based models ~\cite{zeng2005knowledge,saravanan2009improving} and Natural Language Processing (NLP) models\cite{shao2020bert,xiao2021lawformer,chalkidis2020legal,ma2021retrieving}. For Expert Knowledge-based models, Zeng et al.~\cite{zeng2005knowledge} extend the traditional legal issue elements with a new set of sub-elements for the representation of legal cases.

With the development of deep learning, NLP-based models have achieved great success in legal case retrieval. Shao et al.~\cite{shao2020bert} divide the legal case texts into several paragraphs and use BERT to obtain the similarity between the paragraphs, achieving promising ranking performance. Paheli Bhattacharya et al.~
\cite{bhattacharya2022legal} combine textual and network information for estimating legal case similarity.

In recent years, many researchers have attempted to design pre-training techniques to achieve performance gains in the legal community. For instance, Chalkidis et al.~\cite{chalkidis2020legal} collect extensive English legal text from several fields (e.g., legislation, court cases, contracts) and release LEGAL-BERT. Xiao et al.~\cite{xiao2021lawformer} proposed Lawformer for longer legal text, which can process documents with thousands of tokens. However, neither of them designed pre-training objectives tailed for legal case retrieval. We argue that the potential of pre-trained models has not been fully exploited.

\section{Background and Preliminaries}
In this section, we introduce the problem definition of legal case retrieval and preliminary knowledge about legal case documents.

\subsection{Problem Formulation}

Legal case retrieval task refers to finding cases relevant to a given query case from the candidate case set. Formally, given a query case $q$, and a set of candidate cases $\mathbf{D}$, the task is to retrieve top-$k$ relevant cases $\mathbf{D}_q = d_{1}^*, d_{2}^*,......, d_{k}^*$  from a large candidate pool. 

For simplicity, we focus on a dual encoder architecture that has been widely used for retrieval tasks ~\cite{zhan2020repbert,karpukhin2020dense}. This architecture consists of a query encoder and a document encoder, where the primary goal is to map the text to a high-dimensional embedding $h_q$ and $h_d$. Here, the document encoder and the query encoder are typically implemented or initialized with a pre-trained language model which outputs an embedding representation (e.g., the embedding of the [CLS] token) by taking the raw text as inputs. Then, the dot product is used to compute the semantic relevance scores of $q$ and $d$:
\begin{equation}\label{eqn-1} 
  S(q,d)=h_q \cdot h_d
\end{equation}

As illustrated in Figure ~\ref{structure}, a candidate case consists of Fact $F$,  Reasoning $R$, Decision $D$ and etc. In most practical legal case retrieval scenarios, $q$ only consists of the basic Fact $F$, while candidates are complete case documents. Particularly, given the basic facts of a query case, lawyers or judges search with the legal case retrieval system to find relevant cases to the query case so that they can better accomplish their downstream tasks such as judicial judgments. In our work, we follow the above settings and assume that queries are the basic fact of legal case documents.



\begin{figure}[t]
\centerline{\includegraphics[width=\columnwidth]{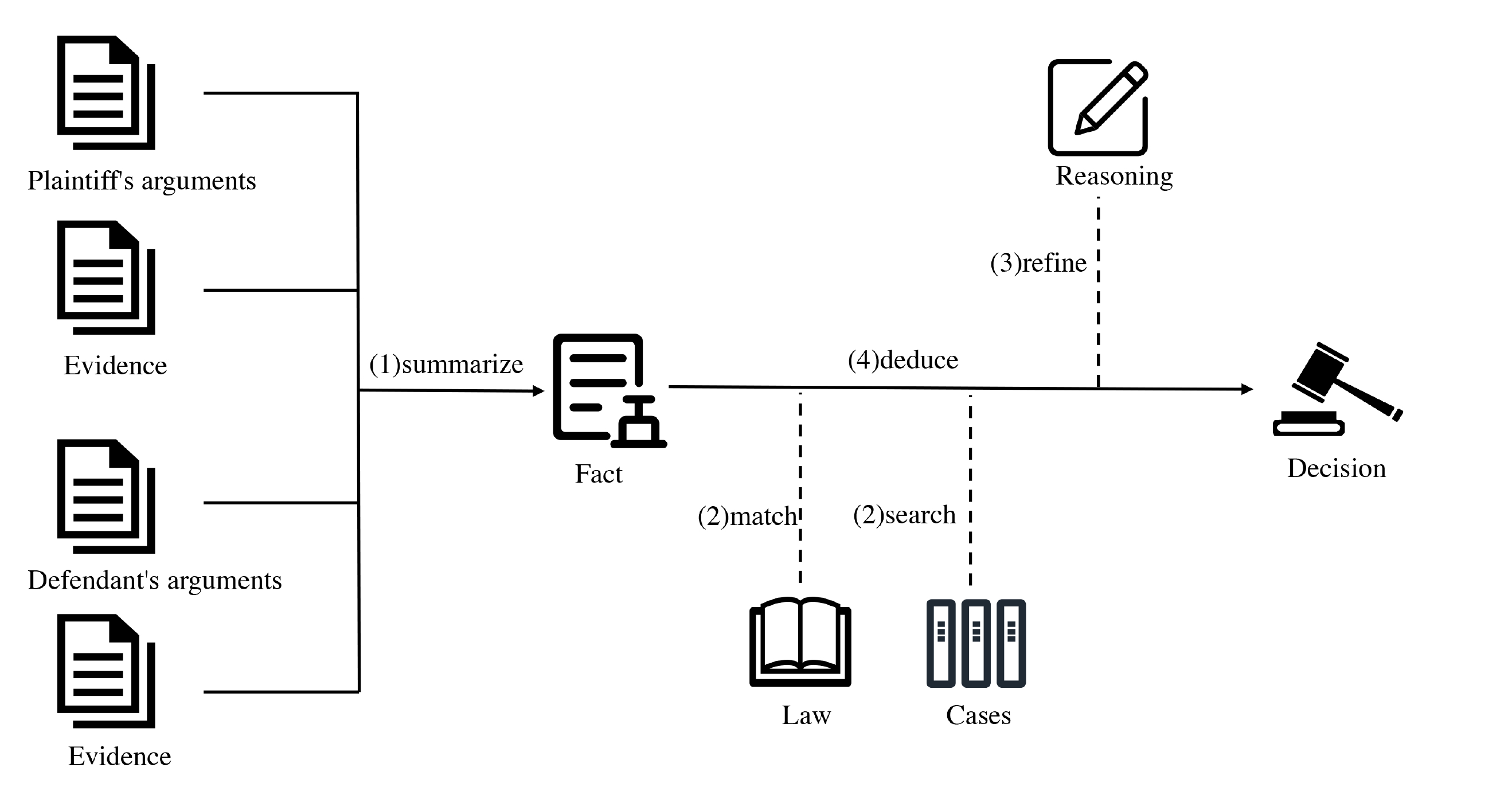}}
\vspace{-0.15in}
\caption{The process of writing a legal document. The process of searching for relevant cases occurs after Fact are obtained. There is significant legal knowledge of judges in the Reasoning and Decision sections.}
\vspace{-5mm}
\label{legal}
\end{figure}

\subsection{Preliminary}
In contrast to documents in open-domain retrieval tasks like Web search, legal case documents usually have clearer yet more complicated structures. Specifically, in countries with a Case Law system, such as the United States, legal case documents generally consist of Procedure, Fact, Reasoning, Decision, and Tail. The Procedure section introduces the parties' information and procedural posture.
The Fact section is a description of the parties' arguments, evidence, and basic events. The Reasoning component is the process where the court selects the rules and applies them to the facts.
In Reasoning, the judge explains the reasons for the application of the rules. In other words, the events that are relevant to the application of the rules, i.e., the key legal elements, are repeatedly mentioned in this section.
The Decision section is the specific response given by the court to the legal dispute based on the key facts of the case. The Reasoning section and the Fact section are the basis of the court's decision. The Tails section introduces the basic information about the court, the judge, etc.

In countries with a Statute Law system, e.g., China, legal case documents have the same structure but are not divided into different sections explicitly. The structure of the Chinese legal case documents is implicitly conveyed by text formatting. For instance, ``after identification" is usually the beginning of the Fact section, and ``The court held that" is followed by the Reasoning section. Despite the differences in legal systems and provisions between countries, these sections are the common foundation of a complete legal case document.

For trials in practice, the process of writing a legal document is illustrated in Figure \ref{legal}. Firstly, The plaintiff's arguments, the defendant's arguments, and their evidence are presented to the court. The judges summarize them to form the Fact part. Then, the judges identify the law that matches the Fact and the cases that are relevant to the Fact. After that, the judges extract the key legal elements and explain the application of the law according to these elements to generate the Reasoning section. Finally, based on their understanding of the cases and legal knowledge, the judges make the final judicial decision, such as the cause of action, the term of penalty, the amount of compensation, etc.

In the above process, important information in the Fact will be carefully analyzed in the Reasoning section, which affects the final decision. This logical connection between Fact, Reasoning, and Decision is important for the understanding and modeling of legal case documents, which inspires us to propose a structure-aware pre-trained language model for legal case retrieval. More details about our proposed method are described in the next section.

\begin{figure}[t]
\centering
\includegraphics[width=\columnwidth]{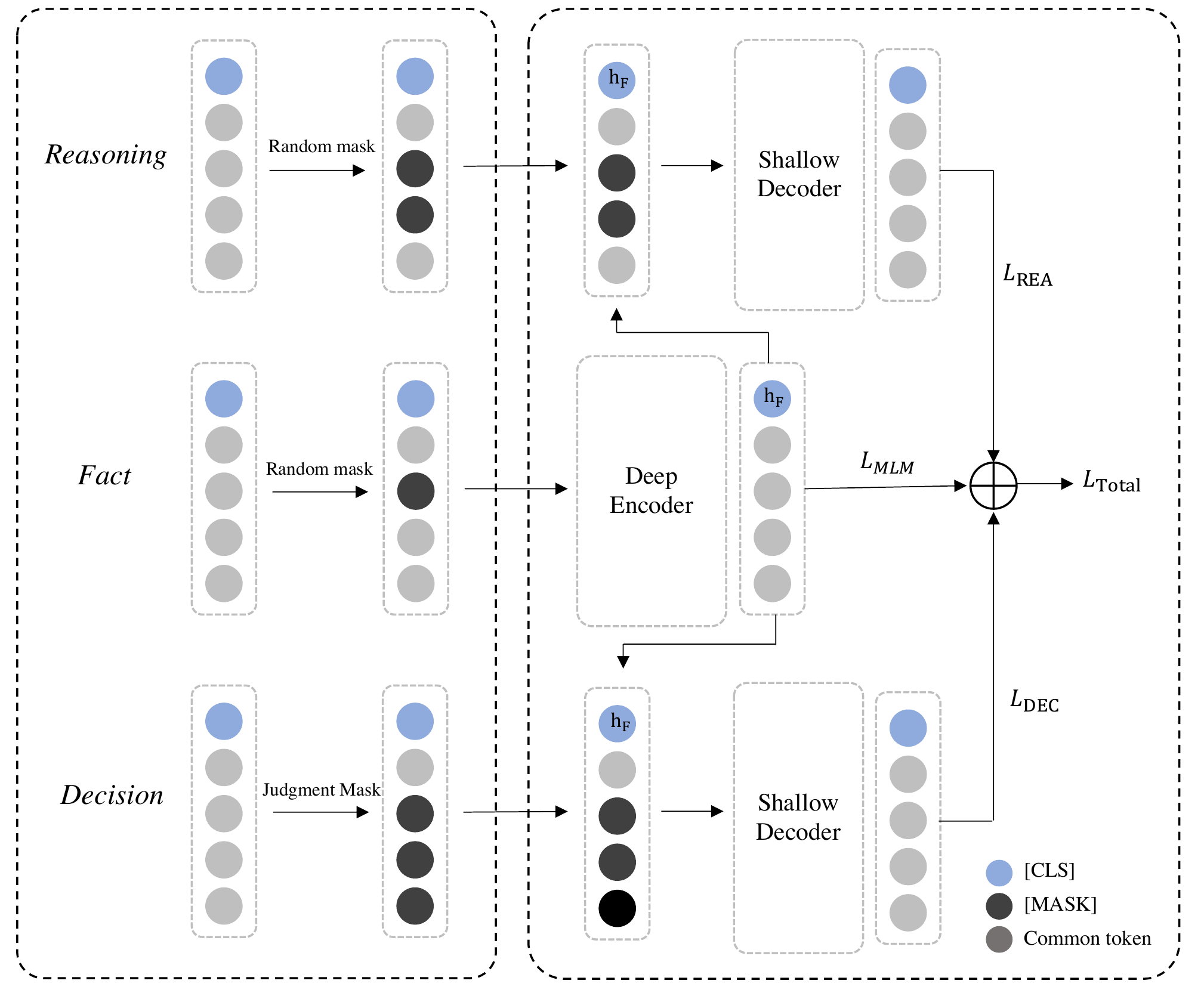}
\vspace{-0.15in}
\caption{The model design for SAILER, which consists of a deep encoder and two shallow decoders. The Reasoning and Decision section are aggressively masked, joined with the Fact embedding to reconstruct the key legal elements and the judgment results. }
\vspace{-5mm}
\label{model}
\end{figure}

\section{Method}
In this section, we describe our structure-aware pre-trained language model for legal case retrieval in detail.

In order to integrate the structural knowledge into the language model, we simulate the writing process of legal case documents and propose the SAILER, which is shown in Figure \ref{model}. SAILER mainly consists of three components, i.e., Fact Encoder, Reasoning Decoder, and Decision Decoder. Specifically, in the pretraining process, a BERT-like encoder is used to generate a vector representation of the Fact, and two shallow decoders are used to reconstruct the text of the Reasoning and Decision sections. Detailed descriptions about the SAILER are as follows.

\subsection{Fact Encoder}
In the Fact Encoder, we first randomly replace some tokens of the Fact with a special token [MASK]. Only a small percentage of tokens is replaced, as the majority of the Fact need to be preserved.

In particular, We define the facts as $F = [f_1,f_2,...,f_n]$ and the masked token set as $m(F)$, where $f$ denotes tokens and $n$ denotes the length of Fact. So the rest tokens can be represented as $F \backslash m(F)$. Then, the masked input $F_{mask}$ is transformed into a sentence vector by the Fact Encoder $\phi_F(\cdot)$. Following the previous work~\cite{devlin2018bert,gao2021condenser}, the final hidden state of the [CLS] token is taken as the representation of the entire sentence:
\begin{equation}\label{eqn-3} 
  h_F = \phi_F(F_{mask}) 
\end{equation}

Similar to Bert, We input the masked facts into the encoder to recover $m(F)$ and compute mask language modeling (MLM) loss as one of our pre-training objectives. Specifically, the MLM loss $L_{MLM}$ is defined as:
\begin{equation}\label{eqn-4} 
  L_{MLM} = -\sum_{x^{'}\in m(F)}\log p(x^{'} | F \backslash m(F))
\end{equation}

\subsection{Reasoning Decoder}
As mentioned above, the Reasoning section contains all the key legal elements of the case. Therefore, we design the Reasoning Decoder to model the logical relationships between Reasoning and Fact, aiming to improve the alignment between the focus of the encoder and the underlining key legal elements in case documents.

Specifically, we construct the Reasoning Decoder through polluting the original text of the Reasoning section $R = [r_1,r_2,...,r_n]$. We split the Reasoning text into tokens and randomly select a subset $m(R)$ for masking. An aggressive mask ratio (30\%-60\%) is adopted for language reconstruction. In addition, the dense vector $h_F$ replaces the original [CLS] vector which is commonly used as the beginning of decoder inputs. The whole decoder input is formalized as:
\begin{equation}\label{eqn-5} 
  [h_F, e_{r_1}+p_{r_1},...,e_{r_n}+p_{r_n}];
\end{equation}
where $e_{r_k}$ and $p_{r_k}$ denote the embeddings and extra position embedding of $r_k$. In this way, the learning objective of the Reasoning Decoder is formulated as:
\begin{equation}\label{eqn-6} 
  L_{REA} = -\sum_{x^{'}\in m(R)}\log p(x^{'} | [h_{F},R \backslash m(R)])
\end{equation}
The Reasoning Decoder needs to rely on $h_F$ to recover the masked Reasoning section, which forces the $h_F$ to focus on useful information related to the Reasoning section in the facts. By this design, SAILER enhances attention to the key legal elements and captures the long dependencies between Facts and Reasoning.

\subsection{Decision Decoder}

Legal Judgment Prediction (LJP) aims to predict the judgment results of a case based on the fact description~\cite{dong2021legal}. If two cases have the same charges and law articles, then their dense vectors should be closer in the vector space. In this part, the Legal Judgment Prediction is modeled as a text-to-text task by using Decision Decoder, which helps the vectors have a stronger discriminatory ability.

For Chinese legal cases, the Decision usually consists of the relevant law articles, charges, and terms of penalty. Given a Decision section $D = [d_1,d_2,...,d_n]$, we mask it according to its specified format. Concretely, the Decision is usually formatted as follows:
\emph{According to the Criminal Law of the People's Republic of China $[z_1]$, the defendant committed $[z_2]$ and was sentenced to $[z_3]$}, where $z_1$ denotes the specific law articles, $z_2$ represents the charges, and $z_3$ is the term of penalty. We use $\mathcal{Z}$ to denote all masked tokens.

In some countries where the Decision does not have a specific format, we select a certain percentage of words with high TF-IDF~\cite{aizawa2003information} values to mask. For convenience, we still name the masked tokens as $\mathcal{Z}$.

The Decision Decoder also relies on the dense vector $h_F$ to recover the original decision. Specifically, the input of the Decision Decoder is:
\begin{equation}\label{eqn-7} 
  [h_F, e_{d_1}+p_{d_1},...,e_{d_n}+p_{d_n}];
\end{equation}
where $e_{d_k}$ and $p_{d_k}$ denote the embeddings and extra position embedding of $d_k$. The Decision Decoder is trained with the objective:

\begin{equation}\label{eqn-8} 
  L_{DEC} = -\sum_{x^{'}\in \mathcal{Z}}\log p(x^{'} | [h_{F},D \backslash \mathcal{Z})])
\end{equation}

In short, Decision Decoder cleverly models the Legal Judgment Prediction task to help generate discriminative vectors. 


\subsection{Learning}
During pre-training, we aim at optimizing the following training objectives:
\begin{equation}\label{eqn-9} 
  L_{Total} = L_{MLM} + L_{REA} + L_{DEC}
\end{equation}

After pre-training, we drop both decoders and fine-tune the encoder. The purpose of fine-tuning is to make the query closer to related cases in the vector space compared to the irrelevant ones. Thus, given a query case $q$, let $d^{+}$ and $d^{-}$ be relevant and negative cases, the loss function $L$ is formulated as follows:

\begin{equation}\label{eqn-10} 
  L(q,d^+,d^-_{1},...,d^-_{n}) =
-\log_{}{    \frac{exp(s(q,d^+))}{exp(s(q,d^+))+\sum_{j=1}^nexp(s(q,d^-_j))}}
\end{equation}
Following previous work~\cite{qu2021rocketqa,zhan2021optimizing}, the negative samples $d^{-}$ are BM25 hard negatives. During training, it is computationally expensive to sample many negative samples for each query case. Therefore, in-batch negatives ~\cite{karpukhin2020dense} are adopted for contrastive learning which to take full advantage of negative samples in the same batch. Specifically, if there are $B$ queries in a batch, each with one positive and $N$ negative examples, then we can obtain at most $(B-1)*(N+1)$ in-batch negative examples for each query during training.

\begin{table*}[ht]
\vspace{-5mm}
\caption{Statistics of benchmark datasets.}
\begin{tabular}{@{}ccccc@{}}
\toprule
DATASETS                         & LeCaRD  & CAIL2022-LCR & COLIEE2020 & COLIEE2021 \\ \midrule
Language                      & chinese & chinese      & english    & english    \\
\# Train query case              & -       & -            & 520        & 650        \\
\# Train candidate cases/query   & -       & -            & 200        & 4,415       \\
\# Test query case               & 107     & 130          & 130        & 250        \\
\# Test candidate cases/query    & 100     & 100          & 200        & 4,415       \\
Avg. length per case document & 8,275    & 2,707         & 3,232       & 1,274       \\
\# Avg. relevant cases per query & 10.33   & 11.53        & 5.15       & 4.73       \\ \bottomrule
\end{tabular}
\label{statistics}
\vspace{-4mm}
\end{table*}

\section{Experiment}
In this section, we first introduce our experimental setup, including datasets and metrics, baselines, and implementation details. Then, we report experimental results to demonstrate the effectiveness of SAILER.
\subsection{Datasets and Metrics}

We conduct our experiments on four legal case retrieval benchmarks.
The statistics are shown in Table \ref{statistics}.

\begin{itemize}[leftmargin=*]
    \item[-] \textbf{LeCaRD}~\cite{ma2021lecard} is the first criminal case retrieval dataset under the Chinese law system. Queries in LeCaRD only contain the fact paragraph while the candidate documents are the entire case. There is a pool of 100 candidates for each query. 
    \item[-] \textbf{CAIL2022-LCR} is the official testing set provided by CAIL2022~\footnote{\url{http://cail.cipsc.org.cn/task3.html?raceID=3&cail_tag=2022}}, which is organized in the same format as LeCaRD.
    \item[-] \textbf{COLIEE2020}~\cite{rabelo2020coliee} is the official dataset provided by COLIEE2020~\footnote{\url{https://sites.ualberta.ca/~rabelo/COLIEE2020/}}. Each query has 200 candidates. Participants need to re-rank a limited number of cases per query.
    \item[-] \textbf{COLIEE2021}~\cite{rabelo2022overview} is the official dataset provided by COLIEE2021~\footnote{\url{https://sites.ualberta.ca/~rabelo/COLIEE2021/}}. Without a candidate document pool, participants need to identify relevant cases from the entire corpus, making the retrieval task more difficult than that in COLIEE2020.

\end{itemize}


We report the zero-shot performance of all models on LeCaRD and CAIL2022-LCR since the number of annotated queries are highly limited. For COLIEE2020 and COLIEE2021, we directly fine-tune the baselines and SAILER with the combined training sets of the two datasets and report the final performance on the testing sets.  

We follow the evaluation metrics in the competition. For the LeCaRD and CAIL2022-LCR, we report normalized discounted cumulative gain (NDCG), Precision@5, Recall@5, and F1 score. For two COLIEE tasks, we report the mean reciprocal rank (MRR), Precision@5, Recall@5, and F1 score. Specifically, we report R@100 for COLIEE2021.

\subsection{Baselines}

We adopt three groups of baselines for comparison: Traditional Retrieval Models, Generic Pre-trained Models, and Retrieval-oriented Pre-trained Models.
\begin{itemize}[leftmargin=*]
    \item \textbf{Traditional Retrieval Models}
    
    \begin{itemize}
        \item[-] \textbf{BM25}~\cite{robertson2009probabilistic} is a highly effective sparse retriever based on exact word matching.
        \item[-] \textbf{QL}~\cite{zhai2008statistical} is another representative traditional retrieval model based on dirichlet smoothing.
    \end{itemize}  

    \item \textbf{Generic Pre-trained Models} 
    
   \begin{itemize}
        \item[-] \textbf{BERT}~\cite{devlin2018bert} is a multi-layer transformer encoder, which is a strong baseline in ad-hoc retrieval tasks.
        \item[-] \textbf{BERT\_xs} ~\footnote{\url{http://zoo.thunlp.org/}} is a criminal law-specific BERT pre-trained with 663 million Chinese criminal judgments.
        \item[-] \textbf{RoBERTa}~\cite{liu2019roberta} is an enhanced version of BERT with substantially enlarged training data. Compared with BERT, RoBERTa is pre-trained only with MLM task.
        \item[-] \textbf{Lawformer}~\cite{xiao2021lawformer} is the first Chinese pre-trained language model in the legal domain, which focuses on processing long documents.
        \item[-] \textbf{LEGAL-BERT}~\cite{chalkidis2020legal} is a pre-trained language model using a large number of English legal documents, achieving state-of-the-art results in multiple tasks.
        
    \end{itemize}
    
    \item \textbf{Retrieval-oriented Pre-trained Models} 
    
    \begin{itemize}
        \item[-] \textbf{Condenser}~\cite{gao2021condenser} is designed for ad-hoc dense retrieval. It utilizes skip connections to aggregate text information into dense vectors.
        \item[-] \textbf{coCondenser}~\cite{gao2021unsupervised} adds an unsupervised corpus-level contrastive loss to Condenser, which can effectively warm up the vector space.
        \item[-] \textbf{SEED}~\cite{lu2021less} theoretically analyzes the insufficiency of the autoencoder architecture for dense retrieval. It uses a weak decoder to enhance the encoder training process.
        \item[-] \textbf{CoT-MAE}~\cite{wu2022contextual} designs efficient data construction approaches to train asymmetric encoder-decoder architectures.
        \item[-] \textbf{RetroMAE}~\cite{liu2022retromae} proposes enhanced decoding which makes it harder to reconstruct text and achieves state-of-the-art performance on ad-hoc retrieval.

    \end{itemize}
\end{itemize}

For traditional models, We use the pyserini toolkit~\footnote{\url{https://github.com/castorini/pyserini}} with the default parameters.
For generic pre-trained models, we directly adopt the open-sourced checkpoint to warm up model parameters. For retrieval-oriented pre-trained models, we use the same legal pre-training corpus as SAILER and pre-train them using the best parameters reported in their paper. All retrieval-oriented pre-trained models are initialized with BERT.

\subsection{Implementation Details}

\subsubsection{Pre-training Procedures}
To construct the pre-training corpus of Chinese legal cases,
we collect tens of millions of case documents from China Judgment Online~\footnote{\url{https://wenshu.court.gov.cn/}}. We divide the case documents into five parts: Procedure, Fact, Reasoning, Decision, and Tail with regular expression matching. Simple cases with facts less than 50 words are filtered. To pre-train the English-based model, we collect a large corpus of case documents from the U.S. federal
and state courts~\footnote{\url{https://case.law/}}. Different from LEGAL-BERT, our pre-trained corpus only contains case documents and excludes legislation and contracts. 

We use chinese-bert-wwm~\footnote{\url{https://huggingface.co/hfl/chinese-bert-wwm}}/bert-base-uncased\footnote{\url{https://huggingface.co/bert-base-uncased}} from Huggingface to initialize the encoder of Chinese/English version of SAILER, respectively. The decoders are randomly initialized transformer layers. The default mask ratio is 0.15 for the encoder and 0.45 for the decoder. We pre-train up to 10 epochs using AdamW~\cite{loshchilov2018fixing} optimizer, with a learning rate of 1e-5, batch size of 72, and linear schedule with warmup ratio 0.1. 
For Decision Decoder, we mask the law articles, charges, and terms of penalty in Chinese cases while masking the words with higher TF-IDF~\cite{aizawa2003information} scores in English cases. The Decision Decoder keeps the same mask ratio as the Reasoning Decoder in the English version of SAILER.


\subsubsection{Fine-tuning Procedures.}
For fine-tuning, given a query, we use BM25 to recall the top 100 related documents from the entire corpus, where irrelevant documents are treated as hard negative samples. We fine-tune up to 20 epochs using the AdamW~\cite{loshchilov2018fixing} optimizer, with a learning rate of 5e-6, batch size of 4, and linear schedule with warmup ratio 0.1. Each batch contains one query and sixteen documents, which means the ratio of positives and hard negatives is 1:15. All the experiments in this work are conducted on 8 NVIDIA Tesla A100 GPUs.

\begin{table*}[t]
  \caption{Zero-shot performance of various baselines. */** denotes that SAILER performs significantly better than baselines at $p < 0.05/0.01$ level using the fisher randomization test. Best method in each column is marked bold. }
  \label{zero-shot}
  \begin{tabular}{lllllllllll}
  \hline
  \multicolumn{1}{l|}{\multirow{2}{*}{Model}} & \multicolumn{5}{c|}{LeCaRD}                                                                                                                                     & \multicolumn{5}{c}{CAIL2022-LCR}                                                                                                                       \\
  \multicolumn{1}{l|}{}                       & \multicolumn{1}{c}{Precision} & \multicolumn{1}{c}{Recall} & \multicolumn{1}{c}{F1\_score} & \multicolumn{1}{c}{NDCG@10} & \multicolumn{1}{l|}{NDCG@30}         & \multicolumn{1}{c}{Precision} & \multicolumn{1}{c}{Recall} & \multicolumn{1}{c}{F1\_score} & \multicolumn{1}{c}{NDCG@10} & \multicolumn{1}{c}{NDCG@30} \\ \hline \hline 
  \multicolumn{11}{c}{\textbf{Traditional Retrieval Models}}                                                                                                                                                                                                                                                                                                             \\ \hline
  \multicolumn{1}{l|}{BM25}                   & 0.8916                        & 0.1748                     & 0.2922                        & 0.7115**                    & \multicolumn{1}{l|}{0.8172*}         & 0.8477**                        & 0.2018**                     & 0.3259**                        & 0.7303**                    & 0.8304*                      \\
  \multicolumn{1}{l|}{QL}                   & 0.8897                        & 0.1737                     & 0.2906                        & 0.7157**                     & \multicolumn{1}{l|}{0.8373}          & 0.8538**             & 0.2004**                     & 0.3246**                        & 0.7535**                      & 0.8545             \\ \hline \hline 
  \multicolumn{11}{c}{\textbf{Generic Pre-trained Models}}                                                                                                                                                                                                                                                                                                               \\ \hline
  \multicolumn{1}{l|}{Chinese BERT}           & 0.6654**                      & 0.1263**                   & 0.2123**                      & 0.5252**                    & \multicolumn{1}{l|}{0.5374**}        & 0.6600**                      & 0.1487**                   & 0.2427**                      & 0.5604**                    & 0.5618**                    \\
  \multicolumn{1}{l|}{Chinese RoBERTa}        & 0.8841                        & 0.1778                     & 0.2960                        & 0.7438**                      & \multicolumn{1}{l|}{0.7897**}        & \textbf{0.9046}                       & 0.2180                   & 0.3513                       & 0.8043**                    & 0.8518                    \\
  \multicolumn{1}{l|}{BERT\_xs}               & 0.7159**                      & 0.1377**                   & 0.2309**                      & 0.5695**                    & \multicolumn{1}{l|}{0.5751**}        & 0.6462**                      & 0.1369**                   & 0.2259**                      & 0.5236**                    & 0.5206**                    \\
  \multicolumn{1}{l|}{Lawformer}              & 0.8056**                      & 0.1552**                   & 0.2603**                      & 0.6216**                    & \multicolumn{1}{l|}{0.6362**}        & 0.7908**                      & 0.1820**                   & 0.2935**                      & 0.6908**                    & 0.6988**                    \\ \hline \hline
  \multicolumn{11}{c}{\textbf{Retrieval-oriented Pre-trained Models}}                                                                                                                                                                                                                                                                                                    \\ \hline
  \multicolumn{1}{l|}{Condenser}              & 0.8280**                      & 0.1632**                    & 0.2727**                       & 0.6469**                    & \multicolumn{1}{l|}{0.7125**}        & 0.8092**                      & 0.1892**                   & 0.3067**                      & 0.6925**                    & 0.7402**                    \\
  \multicolumn{1}{l|}{coCondenser}            & 0.8411**                      & 0.1648**                    & 0.2756**                       & 0.6719**                    & \multicolumn{1}{l|}{0.7404**}        & 0.8138**                      & 0.1931**                   & 0.3121**                       & 0.7063**                    & 0.7607**                    \\
  \multicolumn{1}{l|}{SEED}                   & 0.8411**                      & 0.1575**                   & 0.2653**                      & 0.6721**                    & \multicolumn{1}{l|}{0.7330**}        & 0.8369**                      & 0.1906**                   & 0.3105**                      & 0.7365**                    & 0.7815**                    \\
  \multicolumn{1}{l|}{COT-MAE}                & 0.8467**                      & 0.1567**                   & 0.2644**                      & 0.6815**                    & \multicolumn{1}{l|}{0.7089**}        & 0.8446**                      & 0.1996**                   & 0.3229**                      & 0.7274**                    & 0.7311**                    \\
  \multicolumn{1}{l|}{RetroMAE}               & 0.8505**                      & 0.1675**                    & 0.2799**                       & 0.6876**                    & \multicolumn{1}{l|}{0.7326**}        & 0.8531**                      & 0.1974**                   & 0.3206**                      & 0.7378**                    & 0.7770**                    \\ \hline
  \multicolumn{1}{l|}{SAILER}                 & \textbf{0.9028}               & \textbf{0.1902}            & \textbf{0.3142}               & \textbf{0.7979}             & \multicolumn{1}{l|}{\textbf{0.8485}} & 0.8938               & \textbf{0.2310}            & \textbf{0.3671}               & \textbf{0.8319}             & \textbf{0.8660}             \\ \hline
  \end{tabular}
  \end{table*}

  \begin{table*}[t]
  \caption{Overall fine-tuning performance of various baselines. */** denotes that SAILER performs significantly better than baselines at $p < 0.05/0.01$ level using the fisher randomization test. Best method in each column is marked bold. }
  \label{finetune}
  \begin{tabular}{llllllllllll}
  \hline
  \multicolumn{1}{l|}{\multirow{2}{*}{Model}} & \multicolumn{5}{c|}{COLIEE2020}                                                                                                                               & \multicolumn{6}{c}{COLIEE2021}                                                                                                                                                  \\
  \multicolumn{1}{l|}{}                       & \multicolumn{1}{c}{Precision} & \multicolumn{1}{c}{Recall} & \multicolumn{1}{c}{F1\_score} & \multicolumn{1}{c}{MRR@10} & \multicolumn{1}{c|}{MRR@50}          & \multicolumn{1}{c}{Precision} & \multicolumn{1}{c}{Recall} & \multicolumn{1}{c}{F1\_score} & \multicolumn{1}{c}{MRR@10} & \multicolumn{1}{c}{MRR@50} & \multicolumn{1}{c}{R@100} \\ \hline \hline
  \multicolumn{12}{c}{\textbf{Traditional Retrieval Models}}                                                                                                                                                                                                                                                                                                                                      \\ \hline
  \multicolumn{1}{l|}{BM25}                   & 0.4754**                      & 0.5721**                   & 0.5192**                      & 0.7875**                   & \multicolumn{1}{l|}{0.7907**}        & 0.0760**                       & 0.1521*                     & 0.1014**                        & 0.0893**                   & 0.1017**                   & 0.4671**                  \\
  \multicolumn{1}{l|}{QL}                   & 0.4554**                      & 0.5506**                   & 0.4985**                      & 0.7906**                   & \multicolumn{1}{l|}{0.7934**}        & 0.0760**                       & 0.1369**                   & 0.0977**                      & 0.1257*                    & 0.1359**                   & 0.5222                    \\ \hline \hline
  \multicolumn{12}{c}{\textbf{Generic Pre-trained Models}}                                                                                                                                                                                                                                                                                                                                        \\ \hline
  \multicolumn{1}{l|}{BERT}                   & 0.4542**                      & 0.5588**                   & 0.5011**                      & 0.7923**                   & \multicolumn{1}{l|}{0.7948**}        & 0.0687**                      & 0.1254**                   & 0.0882**                      & 0.1063**                   & 0.1194**                   & 0.4504**                  \\
  \multicolumn{1}{l|}{RoBERTa}                & 0.4639**                      & 0.5862**                   & 0.5155**                      & 0.7613**                   & \multicolumn{1}{l|}{0.7635**}        & 0.0728**                      & 0.1288**                   & 0.0930**                      & 0.1200**                   & 0.1340**                   & 0.5001*                   \\
  \multicolumn{1}{l|}{LEGAL-BERT}            & 0.4262**                      & 0.5544**                   & 0.4817**                      & 0.7571**                   & \multicolumn{1}{l|}{0.7594**}        & 0.0704**                      & 0.1205**                   & 0.0888**                      & 0.0973**                   & 0.1063**                   & 0.3783**                  \\ \hline \hline
  \multicolumn{12}{c}{\textbf{Retrieval-oriented Pre-trained Models}}                                                                                                                                                                                                                                                                                                                             \\ \hline
  \multicolumn{1}{l|}{Condenser}              & 0.4862**                      & 0.6127**                   & 0.5421**                      & 0.8198**                   & \multicolumn{1}{l|}{0.8213**}        & 0.0832*                        & 0.1505*                     & 0.1072*                        & 0.1245*                    & 0.1376**                   & 0.4937**                   \\
  \multicolumn{1}{l|}{coCondenser}            & 0.5000**                      & 0.6287**                   & 0.5570**                      & 0.8337*                    & \multicolumn{1}{l|}{0.8347*}         & 0.0896                        & 0.1559*                     & 0.1138                        & 0.1338*                    & 0.1444*                    & 0.4985**                   \\
  \multicolumn{1}{l|}{SEED}                   & 0.5308                        & 0.6952                     & 0.6019                        & 0.8683                     & \multicolumn{1}{l|}{0.8699}          & 0.0944                        & 0.1731                     & 0.1221                        & 0.1370                     & 0.1473                     & 0.4979**                   \\
  \multicolumn{1}{l|}{CoT-MAE}                & 0.4831**                      & 0.6126**                   & 0.5401**                      & 0.8184**                   & \multicolumn{1}{l|}{0.8192**}        & 0.0888                        & 0.1548*                     & 0.1129                        & 0.1140**                   & 0.1238**                   & 0.4392**                  \\
  \multicolumn{1}{l|}{RetroMAE}               & 0.5354                        & 0.6909                     & 0.6033                        & 0.8669                     & \multicolumn{1}{l|}{0.8682}          & 0.0842*                        & 0.1455**                    & 0.1065*                        & 0.1253*                    & 0.1357**                   & 0.4727**                  \\ \hline
  \multicolumn{1}{l|}{SAILER}                 & \textbf{0.5446}               & \textbf{0.7152}            & \textbf{0.6164}               & \textbf{0.8823}            & \multicolumn{1}{l|}{\textbf{0.8831}} & \textbf{0.1040}               & \textbf{0.1855}            & \textbf{0.1332}               & \textbf{0.1501}            & \textbf{0.1610}            & \textbf{0.5298}           \\ \hline
  \end{tabular}
  \end{table*}

\subsection{Experiment Result}

\subsubsection{Zero-Shot Evaluation}
To verify the effectiveness of SAILER, we conduct zero-shot experiments on two Chinese criminal law datasets. The performance comparisons between SAILER and baselines are shown in Table \ref{zero-shot}. We derive the following observations from the experiment results.
\begin{itemize}[leftmargin=*]
    \item Under the zero resource setting, BM25 and QL provide competitive performance as pre-trained models on legal case retrieval task.
    \item Generic pre-trained models usually perform worse than traditional lexical matching approaches. Without a specific pre-training objective, language models such as BERT are unable to accurately capture the concept of relevance in legal case retrieval.
    \item Retrieval-oriented pre-trained models usually perform better than generic pre-trained models. This suggests that applying retrieval-specific pre-training objectives rather than generic NLP objectives is beneficial for retrieval task.
    \item Surprisingly, Chinese RoBERTa is more effective than models trained with massive criminal law data. Without the next sentence prediction (NSP) task, RoBERTa improves the robustness of the [CLS] embedding. This may indicate that the
    next sentence prediction task is not helpful for improving the retrieval performance of PLMs.
    \item Finally, we can notice that SAILER achieves the best performance in most metrics on both datasets. Without any supervised data, SAILER not only significantly outperforms other pre-trained models, but is also the only model that beats traditional retrieval methods. This observation indicates the effectiveness of incorporating knowledge from the case structure into the pre-training process, revealing the great potential of SAILER in scenarios with no supervised data available.
\end{itemize}

\begin{table}[t]
\caption{Ablation study on LeCaRD under zero-shot setting. Best results are marked bold.}
\begin{tabular}{llccc}
\hline
\multicolumn{2}{l}{Method}                & NDCG@10         & NDCG@20         & NDCG@30         \\ \hline
\multicolumn{2}{l}{SAILER}                & \textbf{0.7979} & \textbf{0.8226} & \textbf{0.8485} \\
\multicolumn{2}{l}{SAILE\_share}          & 0.7869          & 0.8009          & 0.8300          \\
\multicolumn{2}{l}{w/o Decision Decoder}  & 0.7226          & 0.7442          & 0.7884          \\
\multicolumn{2}{l}{w/o Reasoning Decoder} & 0.7385          & 0.7607          & 0.7956          \\
\multicolumn{2}{l}{w/o Both}              & 0.6479          & 0.6684          & 0.6867          \\ \hline
\end{tabular}
\vspace{-3mm}
\label{ablation study}
\end{table}

\subsubsection{Fine-tuning Evaluation}
As shown in Table \ref{finetune}, we compared SAILER with various baselines under the training data. For a fair comparison, different pre-trained models leverage the same fine-tuning data and the same size of hyperparameters. From the experimental results, we have the following findings:

\begin{itemize}[leftmargin=*]
    \item Under the fine-tuning setting, traditional retrieval models (e.g. BM25, QL) still perform well on both datasets.
    \item With the guidance of annotated data, pre-trained models are further improved. However, they are still inferior to traditional retrieval methods in some aspects, e.g., QL outperforms most baselines on the R@100 metric of the COLIEE2021.
    
    \item Overall, SAILER achieves improvements over all the baselines. SAILER takes full advantage of the intrinsic structure of legal cases by inferring expert knowledge through contextual tokens. Without complex enhanced decoding, it can already outperform RetroMAE. On COLIEE2021, SAILER is the only pre-trained model that exceeds QL at Recall@100 metric, indicating it can better distinguish confused cases in a large corpus.

\end{itemize}

As the number of test cases is limited, actually the previous work has not conducted significance tests on COLIEE datasets~\cite{shao2020bert,althammer2021dossier}. Although the improvement of SAILER over the SEED and RetroMAE models is not very significant on the COLIEE2020 dataset, SAILER still achieves state-of-the-art metric values. Generally, PLMs with the same parameter size show close performance after fine-tuning. Therefore, a non-significant improvement on performance can also indicate the effectiveness of our approach.

\begin{table}[t]
\caption{Impact of mask rate $(\%)$ on LaCaRD. Best results are marked bold.}
\label{mask rate}
\begin{tabular}{cc|ccc}
\hline
\multicolumn{1}{l}{Enc} & \multicolumn{1}{l|}{Dec} & \multicolumn{1}{l}{NDCG@10} & \multicolumn{1}{l}{NDCG@20} & \multicolumn{1}{l}{NDCG@30} \\ \hline
0                       & 15                       & 0.7542                      & 0.7746                      & 0.8173                      \\
0                       & 30                       & 0.7685                      & 0.7846                      & 0.8202                      \\
0                       & 45                       & 0.7739                      & 0.7905                      & 0.8262                      \\
0                       & 60                       & 0.7740                      & 0.7858                      & 0.8285                      \\
15                      & 15                       & 0.7778                      & 0.7950                       & 0.8275                      \\
15                      & 30                       & 0.7835                      & 0.7995                      & 0.8343                      \\
15                      & 45                       & \textbf{0.7979}                      & \textbf{0.8226}                      & \textbf{0.8485}                      \\
15                      & 60                       & 0.7865                      & 0.8052                      & 0.8412                      \\
30                      & 15                       & 0.7946                      & 0.8048                      & 0.8361                      \\
30                      & 30                       & 0.7952                      & 0.8089                      & 0.8378                      \\
30                      & 45                       & 0.7915                      & 0.8001                      & 0.8409                      \\
30                      & 60                       & 0.7830                      & 0.8052                      & 0.8445                      \\ \hline
\end{tabular}
\vspace{-3mm}
\end{table}

\begin{table}[ht]
\caption{Impact of decoder layer number on LeCaRD. Best results are marked bold.}
\label{decoder layer}
\begin{tabular}{@{}lccccc@{}}
\toprule
Decoder Layer & 1      & 2      & 3      & 4      & 5      \\ \midrule
NDCG@10       & \textbf{0.7979} & 0.7895 & 0.7806 & 0.7776 & 0.7769 \\
NDCG@20       & \textbf{0.8226} & 0.8081 & 0.7946 & 0.7889 & 0.7837 \\
NDCG@30       & \textbf{0.8485} & 0.8405 & 0.8341 & 0.8253 & 0.8212 \\ \bottomrule
\end{tabular}
\vspace{-3mm}
\end{table}

\subsection{Ablation Studies}
To better illustrate the effectiveness of the model design and pretraining tasks, we further conduct ablation studies on LeCaRD in zero-shot setting. Table \ref{ablation study} shows the impact of different pre-training objectives. SAILER\_{share} is a variant of SAILER, where reasoning and decision sections are reconstructed by a shared decoder. We can observe that sharing a single decoder causes slight performance degradation, which confirms the effectiveness of our multi-decoder architecture. In addition, we remove the Reasoning Decoder and Decision Decoder respectively, both of which cause a significant decrease in performance.
Compared with the removal of the Decision Decoder, the removal of the Reasoning Decoder lead to less performance degradation. One possible reason is that the random mask we used in the Reasoning Decoder could not effectively mask all the key legal elements and thus provides limited guidance for SAILER. How to determine the position of key legal elements in the Reasoning section is one of our future works. Finally, after removing all decoders, the performance decreases dramatically. The above experiments verify the effectiveness of our pre-training objectives.





\subsection{Hyperparameter Analysis}
\subsubsection{Impact of Mask Ratio}
In this section, we explore the impact of different masking rates on SAILER. We conducted experiments on LeCaRD where the encoder masking ratio varies from 0 to 0.30 and the decoder masking ratio is increased from 0.15 to 0.60. The results are shown in Table \ref{mask rate}. There are several interesting findings. (1) The increase of the decoder mask ratio will make the decoding process more difficult and the model performance will increase to some extent. This growth will become unstable when the decoder masking rate is higher than 0.45. We think that excessively difficult decoding may be disadvantageous to the learning of the model. (2) An appropriate encoder mask ratio helps improve the performance. However, SAILER's performance decreases slightly when the encoder masking ratio is too high. This is because a too-large masking ratio will prevent the generation of high-quality sentence embeddings, considering that most of the useful information of the input sentence will be discarded under this circumstance. (3) Overall, SAILER can perform well over a wide range of mask ratios, indicating its robustness. When the encoder mask ratio is 0.15 and the decoder mask ratio is 0.45, SAILER achieves the best performance.

\subsubsection{Impact of Decoder Layer Number}
We further explore the impact of the number of decoder layers on the performance. We keep the same number of layers for the Reasoning Decoder and the Decision Decoder in our experiments. The experimental results are displayed in Table \ref{decoder layer}. With the increasing number of decoder layers, the performance of SAILER decreases, which is similar to the finding of SEED~\cite{lu2021less}. Overall, SAILER's performance with regard to decoder layers is quite robust.

\begin{figure}[t]
\centerline{\includegraphics[width=\columnwidth]{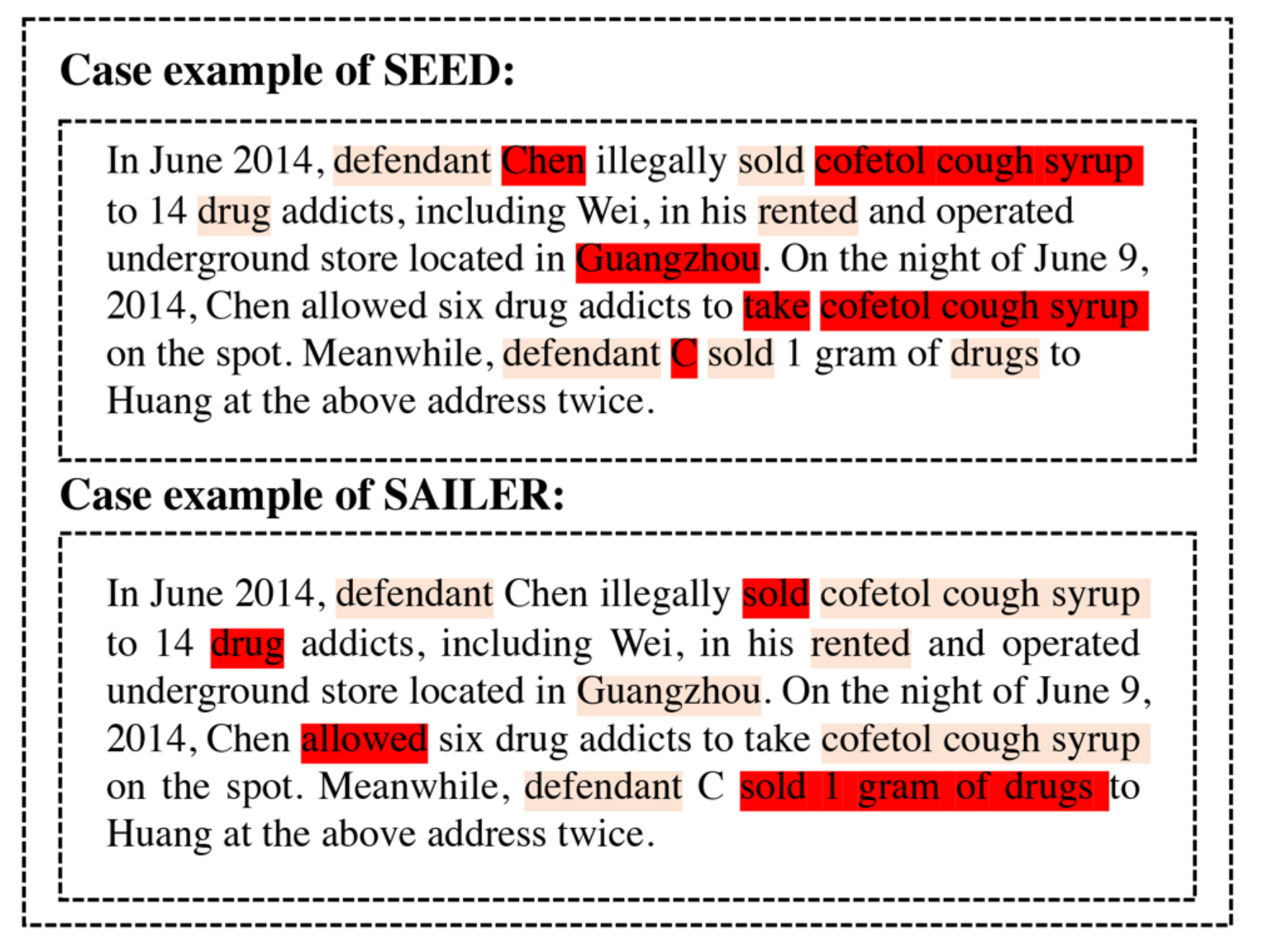}}
\vspace{-4mm}
\caption{Comparison of attention weight visualization for SEED and SAILER. A darker color means a higher attention.}
\label{case study}
\end{figure}

\subsection{Case Study}
To further analyze the retrieval mechanism of SAILER, we visualize the attention weights of words in Figure \ref{case study}, where the darker a word is, the higher the attention weight it gets. 
For a fair comparison, we select SEED, which reconstructs the original text using the same decoding mechanism as SAILER, to conduct the case study. The attention distribution of SEED and SAILER has many differences. It can be observed that SEED pays the most attention to cofetol cough syrup (name of a drug),  Guangzhou (city name), Chen (human name), etc. Whereas SAILER is more concerned with words that affect the final judgment, such as sold, drug, allowed, etc. It is worth mentioning that ``allow" is a keyword in this case, because providing venues for drug users is a typical charge, which means that the defendant allows others to take drugs at his own venue. We can observe that ``allow" is emphasized in SAILER. This shows that SAILER is incorporated with legal knowledge and pays more attention to key legal elements.

\subsection{Visual Analysis}
\begin{figure}[t]
	\centering
	
	\subfigure{
		\begin{minipage}[t]{0.45\columnwidth}
			\centering
			\includegraphics[width=1\linewidth]{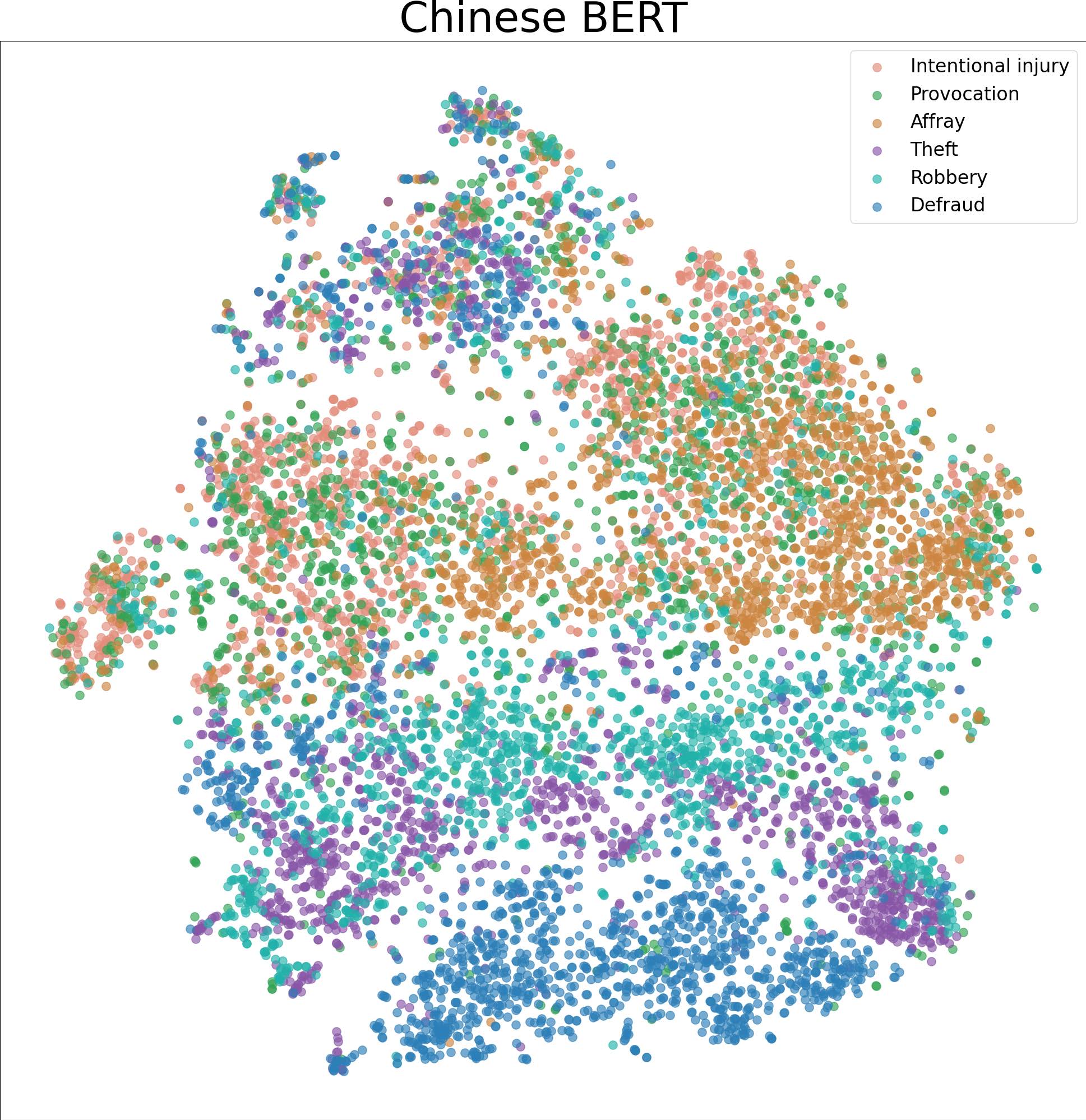}
		\end{minipage}
	}
	\subfigure{
		\begin{minipage}[t]{0.45\linewidth}
			\centering
			\includegraphics[width=1\linewidth]{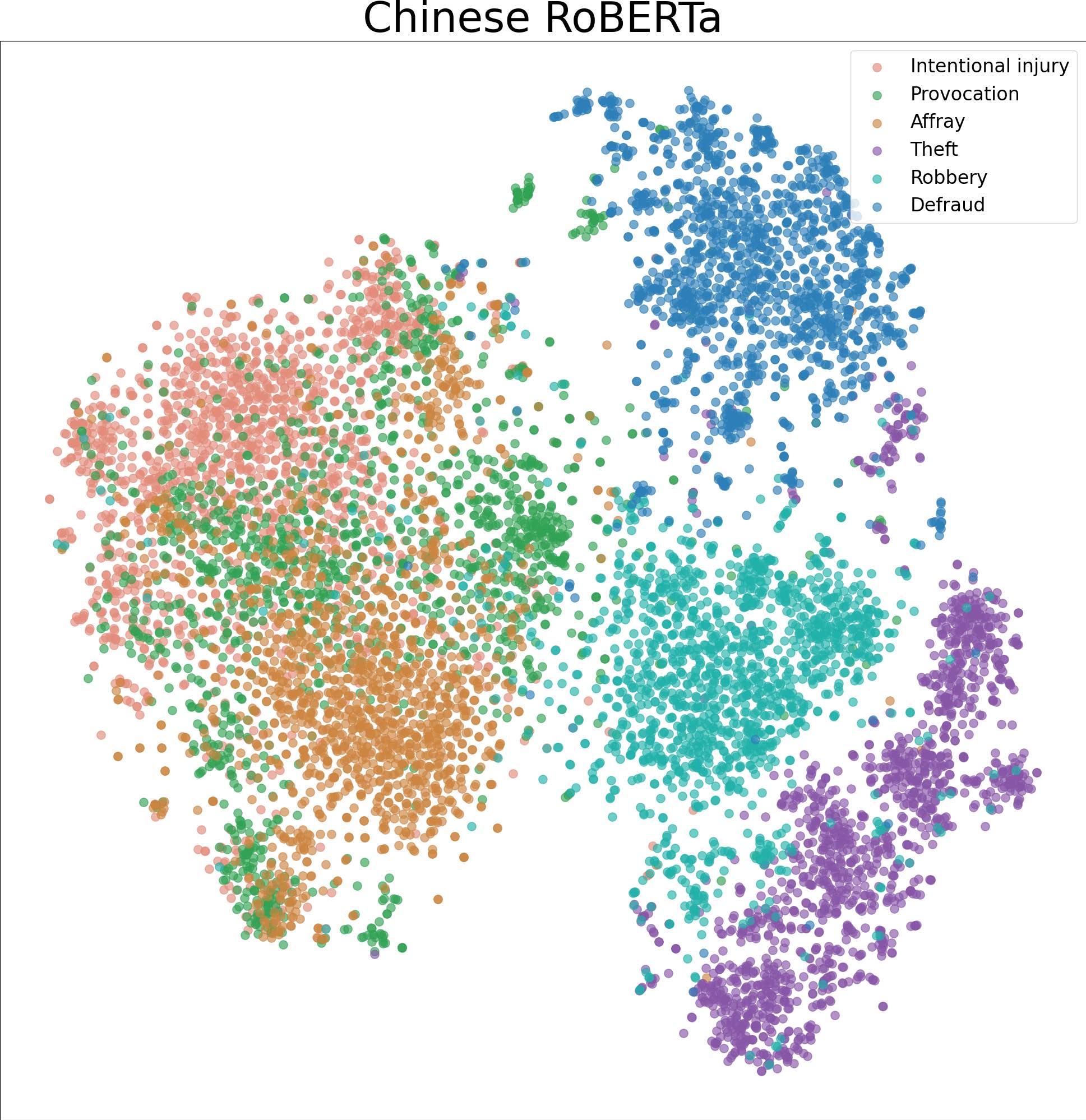}
		\end{minipage}
	}

    \subfigure{
		
		\begin{minipage}[t]{0.45\linewidth}
			\centering
			\includegraphics[width=1\linewidth]{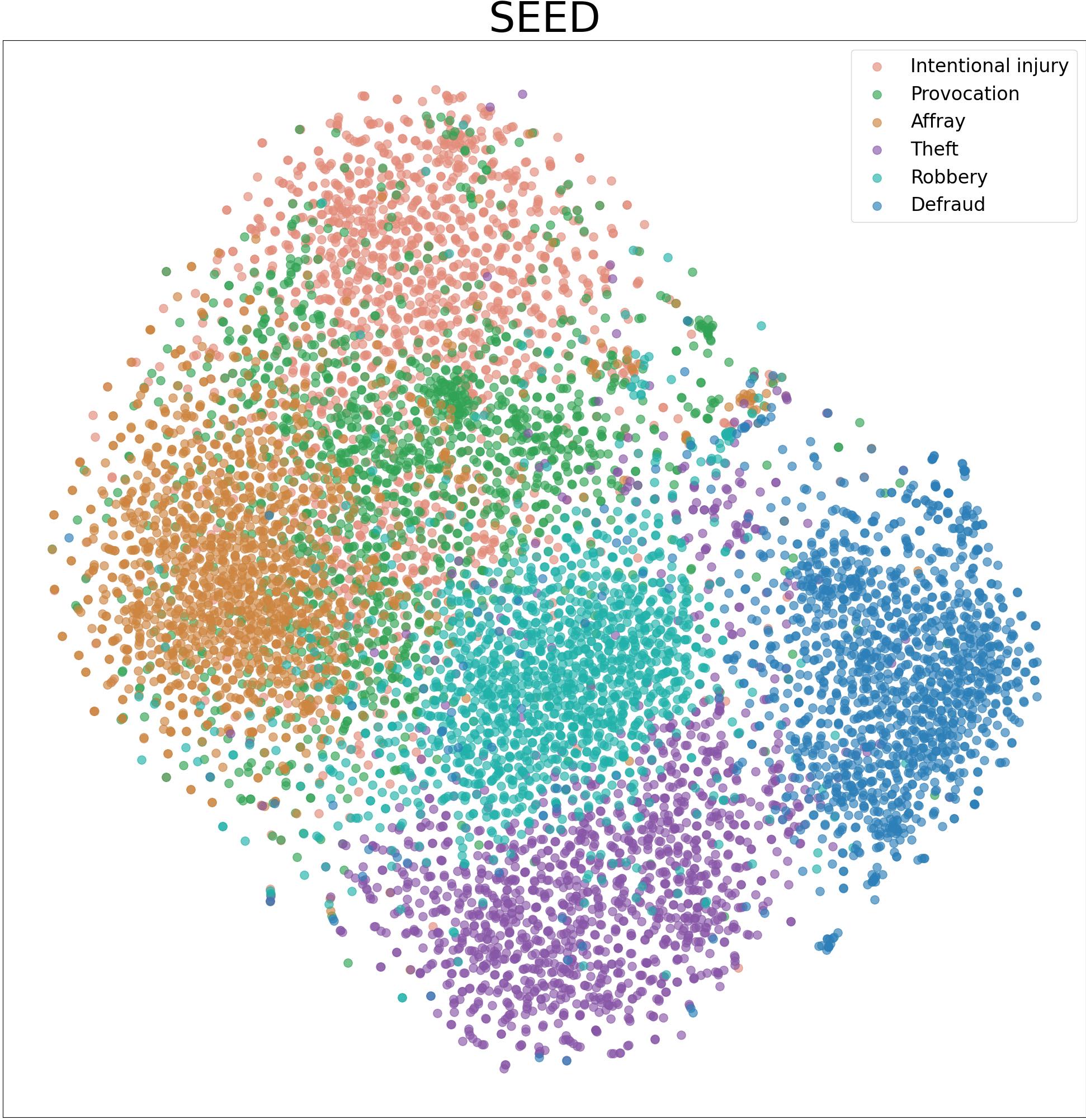}
		\end{minipage}
	}
	\subfigure{
		\begin{minipage}[t]{0.45\linewidth}
			\centering
			\includegraphics[width=1\linewidth]{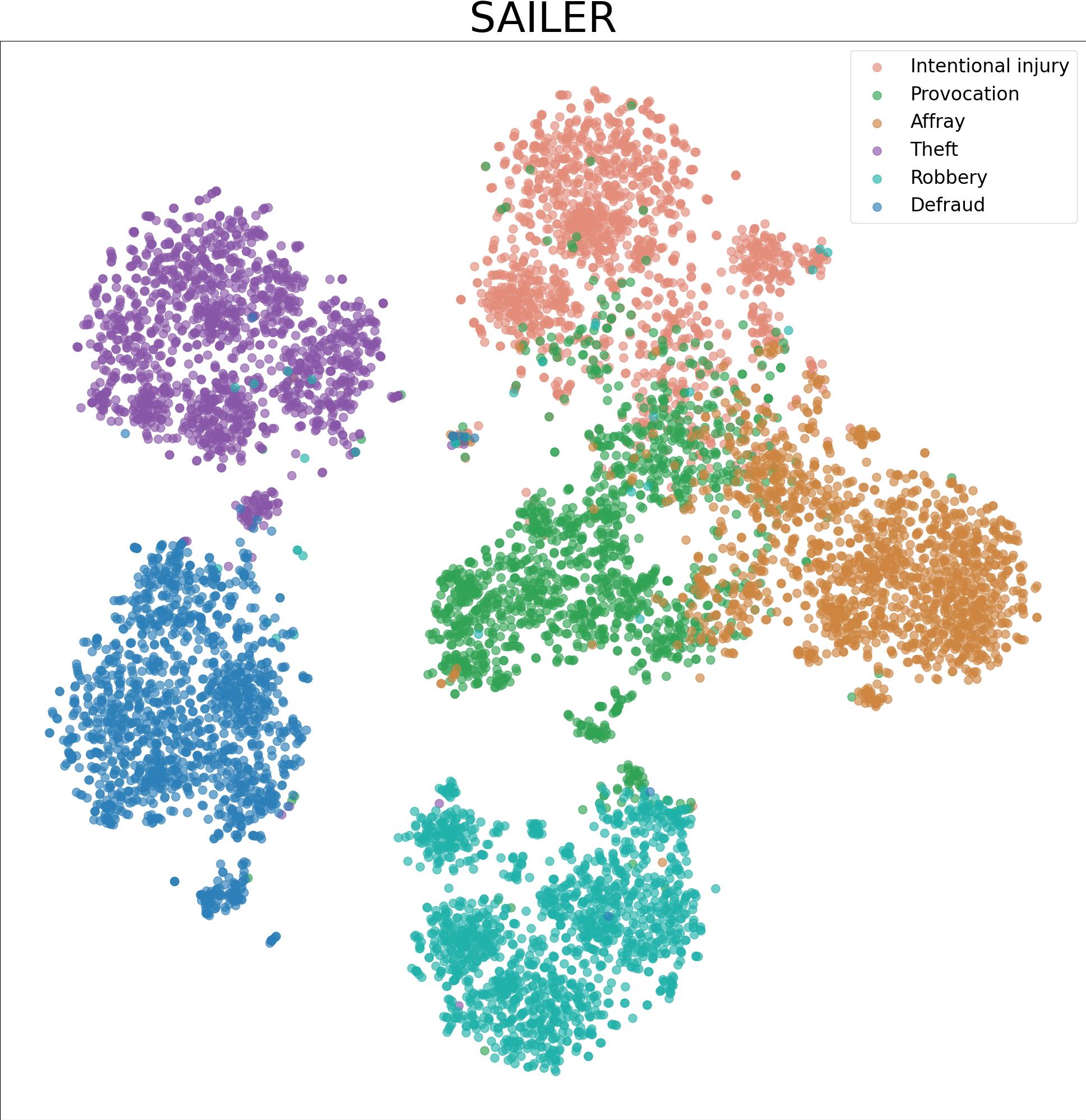}
		\end{minipage}
	}

	\caption{The t-SNE plot of legal cases.}
	\label{vis}
\vspace{-5mm}
\end{figure}

To further explore the discriminative ability of SAILER, we visualize the vector distribution of legal cases by using t-SNE.
Specifically, we selected six charges that are easy to be confused, including Intentional Injury, Provocation, Affray, Theft, Robbery and Defraud. For each charge, a certain number of legal cases were randomly selected from the pre-training corpus. We generate vectors of legal cases using off-the-shelf pre-trained models in zero-shot setting. From Figure ~\ref{vis}, we can observe the powerful discriminative power of SAILER. For Chinse BERT, it always mixes the cases of different charges. Although, Chinese RoBERTa can discriminate three charges of Theft, Robbery and Defraud, it confuses Intentional Injury, Provocation and Affray.
By reconstructing the original text, SEED learns more information and generates vectors in a uniform distribution. However, the boundaries of similar charges are too close, which is not beneficial for legal case retrieval.
In contrast, SAILER can distinguish the vectors of different charges more accurately without any supervision data, indicating that it has learned rich legal knowledge through our pre-training objectives.

\section{Conclusion}
This paper proposes SAILER, a new structure-aware
pre-trained language model for legal case retrieval. The key idea of SAILER is to make full use of the structure relationships of legal case documents for pre-training. By reconstructing key legal elements and judgment results, SAILER generates better legal case representations and owns a stronger discriminative ability. Through extensive experiments on four benchmark legal datasets, SAILER has achieved significant improvements over baselines in both low-resource and full-resource settings. In the future, we would like to explore incorporating more expert knowledge such as legal knowledge graphs and law articles into pre-trained language models for better legal case retrieval.

\begin{acks}
This work is supported by the Natural Science Foundation of China (62002194), Tsinghua University Guoqiang Research Institute, Tsinghua-Tencent Tiangong Institute for Intelligent Computing and the Quan Cheng Laboratory.
\end{acks}

\clearpage
\balance
\bibliographystyle{ACM-Reference-Format}
\bibliography{sample-base.bib}
\end{document}